\documentclass[%
aps,
 pra,%
 amsmath,amssymb,
 reprint,%
 superscriptaddress,
]{revtex4-1}

\usepackage{dcolumn}
\usepackage{graphicx}
\usepackage{lineno}
\usepackage{bm}%
\usepackage[pdftex,colorlinks=true,bookmarks=false,citecolor=blue,urlcolor=blue]{hyperref}
\usepackage{upgreek}
\usepackage{titlesec}

\begin{document}

%%%%%%%%%%%%%%%%%%%%%%%%%%%%%%%%%%%%%%%%%%%%%%%%%%%%%%%%%%%%%%%%%%%%%
%% TITLE SECTION
%%%%%%%%%%%%%%%%%%%%%%%%%%%%%%%%%%%%%%%%%%%%%%%%%%%%%%%%%%%%%%%%%%%%%

\title{\vspace{-15mm}\fontsize{19pt}{10pt}\selectfont\textbf{High visibility time-energy entangled photons from a silicon nanophotonic chip}} % Article title

\author{Steven Rogers}
\altaffiliation{These authors contributed equally to this work}
\affiliation{Department of Physics and Astronomy, University of Rochester, Rochester, NY 14627}
\affiliation{Center for Coherence and Quantum Optics, University of Rochester, Rochester, NY 14627}

\author{Daniel Mulkey}
\altaffiliation{These authors contributed equally to this work}
\affiliation{Center for Coherence and Quantum Optics, University of Rochester, Rochester, NY 14627}
\affiliation{Institute of Optics, University of Rochester, Rochester, NY 14627}

\author{Xiyuan Lu}
\affiliation{Department of Physics and Astronomy, University of Rochester, Rochester, NY 14627}
\affiliation{Center for Coherence and Quantum Optics, University of Rochester, Rochester, NY 14627}

\author{Wei C. Jiang}
\affiliation{Center for Coherence and Quantum Optics, University of Rochester, Rochester, NY 14627}
\affiliation{Institute of Optics, University of Rochester, Rochester, NY 14627}

\author{Qiang Lin}
\email[Electronic mail: ]{qiang.lin@rochester.edu}
\affiliation{Center for Coherence and Quantum Optics, University of Rochester, Rochester, NY 14627}
\affiliation{Institute of Optics, University of Rochester, Rochester, NY 14627}
\affiliation{Department of Electrical and Computer Engineering, University of Rochester, Rochester, NY 14627}

\date{\today}

%----------------------------------------------------------------------------------------

%%%%%%%%%%%%%%%%%%%%%%%%%%%%%%%%%%%%%%%%%%%%%%%%%%%%%%%%%%%%%%%%%%%%%
%% ABSTRACT
%%%%%%%%%%%%%%%%%%%%%%%%%%%%%%%%%%%%%%%%%%%%%%%%%%%%%%%%%%%%%%%%%%%%%

\begin{abstract}
Advances in quantum photonics have shown that chip-scale quantum devices are translating from the realm of basic research to applied technologies. Recent developments in integrated photonic circuits and single photon detectors indicate that the bottleneck for fidelity in quantum photonic processes will ultimately lie with the photon sources. We present and demonstrate a silicon nanophotonic chip capable of emitting telecommunication band photon pairs that exhibit the highest raw degree of time-energy entanglement from a micro/nanoscale source, to date. Biphotons are generated through cavity-enhanced spontaneous four-wave mixing (SFWM) in a high-Q silicon microdisk resonator, wherein the nature of the triply-resonant generation process leads to a dramatic Purcell enhancement, resulting in highly efficient pair creation rates as well as extreme suppression of the photon noise background. The combination of the excellent photon source and a new phase locking technique, allow for the observation of a nearly perfect coincidence visibility of (96.6 $\pm$ 1.1)$\%$, without any background subtraction, at a large pair generation rate of $(4.40\pm0.07) \times 10^5 ~{\rm pairs/s}$.    
\end{abstract}

\maketitle % Insert title

%%%%%%%%%%%%%%%%%%%%%%%%%%%%%%%%%%%%%%%%%%%%%%%%%%%%%%%%%%%%%%%%%%%%%
%% INTRODUCTION
%%%%%%%%%%%%%%%%%%%%%%%%%%%%%%%%%%%%%%%%%%%%%%%%%%%%%%%%%%%%%%%%%%%%%

The field of quantum photonics continues to elucidate the fundamental physics of light-matter interactions and is poised to produce disruptive technologies in the coming years. Quantum photonic based technologies are particularly promising in the areas of quantum communication \cite{Gisin07,Zbinden02,Villoresi15}, quantum simulation \cite{Aspuru-Guzik12,Pan09}, quantum computing \cite{Milburn01,Milburn07,Pittman02,OBrien09,OBrien15}, and quantum metrology \cite{Dowling02,Maccone04}. High visibility quantum interference is essential for the success of many quantum photonic applications \cite{OBrien09,Zeilinger12}. In particular, time-energy entangled photon pairs at telecommunication wavelengths are especially well suited for integration with the current fiber optic infrastructure \cite{Gisin07,Zbinden02}. Recently, there has been significant interest towards creating chip-scale sources of time-energy entangled photons \cite{Ramelow15,Obrien13,Wakabayashi15,Preble15,Grassani15,Xiong15}. Here, we demonstrate a silicon nanophotonic chip that yields the highest raw quantum interference visibility for time-energy entangled photons from a micro/nanoscale source ever reported \cite{Ramelow15,Obrien13,Wakabayashi15,Preble15,Grassani15,Xiong15}. The device was designed to emit photon pairs in the telecommunication wavelength band and fabricated using a CMOS compatible process flow. The nanophotonic chip is capable of producing photon pairs with high spectral brightness, long coherence times, wavelength versatility, and remarkably pure two-photon interference, making it an ideal candidate for future quantum photonic technologies. 

\begin{figure*}[t!]
\begin{center}
\includegraphics[scale=1.1]{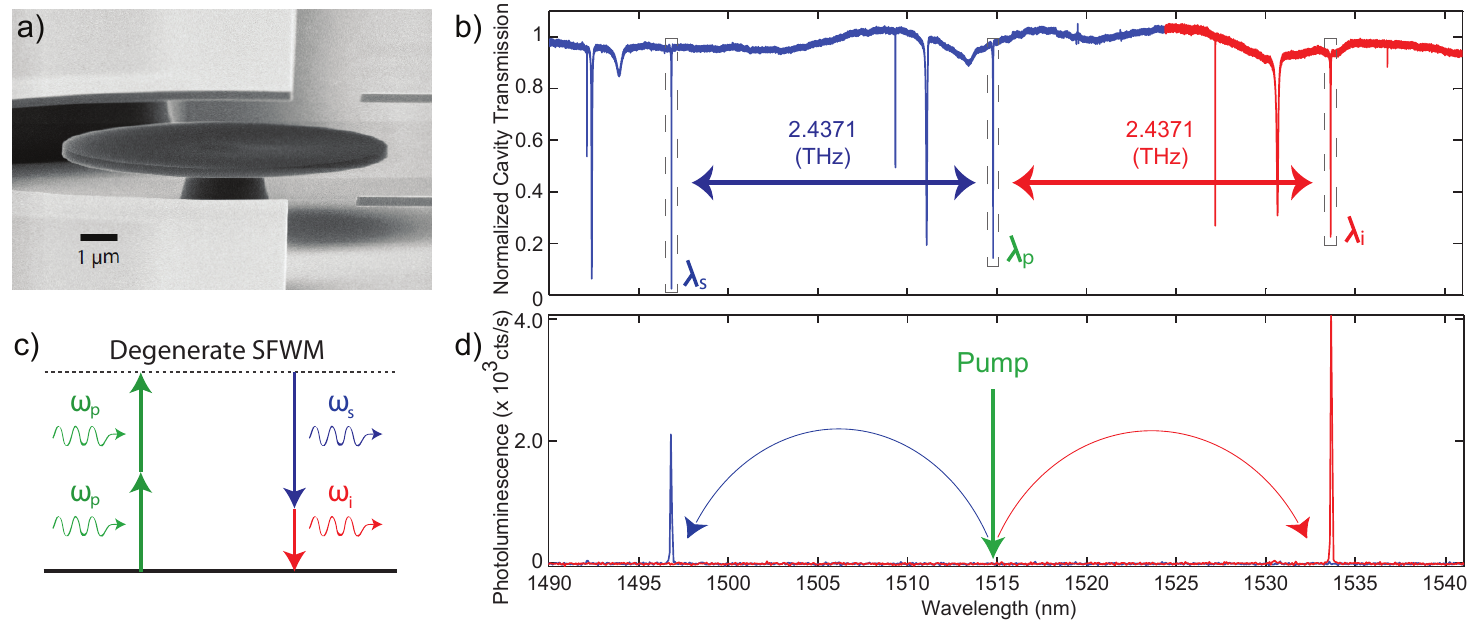}
\caption{ \small Characterization of the silicon microdisk and biphoton generation process. (a) A scanning electron microscopic (SEM) image of the silicon microdisk sitting on a silica pedestal. (b) Normalized cavity transmission trace with boxes indicating the modes used in the spontaneous four-wave mixing (SFWM) process. A verification of frequency matching among the three interacting modes is shown in the inset. (c) Energy diagram depicting the degenerate SFWM process (d) A photoluminescence spectrum with peaks verifying the creation of photons at the signal (blue) and idler (red) wavelengths.  }
\label{Fig1}
\end{center}
\end{figure*}

The device used to generate time-energy entangled photon pairs is a high-Q silicon microdisk, as shown in Fig.~\ref{Fig1}(a), with a radius of approximately 4.5~$\mu$m and a thickness of 260 nm. A normalized cavity transmission trace is shown in Fig.~\ref{Fig1}(b), with dashed boxes indicating the  quasi-transverse-magnetic (quasi-TM) mode family employed for the experiment. Frequency matching among the three cavity modes is achieved by engineering the group-velocity dispersion of the device and is confirmed by the equal mode spacing shown in the inset of Fig.~\ref{Fig1}(b). Consequently, efficient photon generation occurs through cavity-enhanced spontaneous four-wave mixing (SFWM) (see Fig.~\ref{Fig1}(c)), by evanescently coupling pump light, via a tapered optical fiber, into the central resonance at $\lambda_{p}$ = 1514.7 nm, which creates pairs of signal and idler photons at $\lambda_{s}$ = 1496.8 nm and $\lambda_{i}$ = 1534.3 nm, respectively. Moreover, the cavity resonances were observed to exhibit high optical Qs $\sim 0.5\times10^6$ (intrinsic), and a finite-element method simulation indicates a small effective mode volume of $10~{\rm \mu m}^3$. Hence, resonant modes experience a dramatic Purcell enhancement, which transforms the vacuum density of states accessible to the interacting photons and thus enhances both the efficiency and purity of the generated pairs. In addition, the microdisk is suspended above the oxide layer (see Fig.~\ref{Fig1}(a)), so that optical interactions are strongly confined within the single-crystalline silicon, which precludes parasitic Raman photon generation from occurring within the device \cite{Jiang15}. These unique features are clearly evidenced in Fig.~\ref{Fig1}(d), which shows strong photoluminescence peaks at the signal and idler wavelengths, along with a very clean noise floor; properties that are highly desirable for quantum photonic applications \cite{Jiang15,Lu15,Rogers15}.  

The cavity-enhanced SFWM process occurs when pairs of pump photons annihilate to create signal and idler photons \cite{Boyd08} (see Fig.~\ref{Fig1}(c)). The optical parametric process is mediated through the $\chi^{(3)}$ nonlinearity, leading to essentially instantaneous pair creation inside the device, after which signal and idler photons are coupled out of the cavity through the same tapered optical fiber used for delivery. Detailed analysis shows that the probability density to emit pairs of photons from the microresonator, into the optical fiber, at times $t_s$ and $t_i$, is given by \cite{Jiang15} 
\begin{eqnarray}
	p_c(t_s,t_i) = \frac{\Gamma_{es} \Gamma_{ei}}{\bar \Gamma^2}(g N_p)^2 \emph{e}^{-\Gamma_{tj} |t_s -t_i|}, \label{PairProb}
\end{eqnarray}
where $\Gamma_{es}$ and $\Gamma_{ei}$ are the external photon coupling rates at the signal and idler frequencies, respectively, and $\Gamma_{tj}$ (j = s,i) are the photon decay rates of the loaded cavity.  $N_p$ is the number of pump photons inside the cavity. The exponent in equation (\ref{PairProb}) is $\Gamma_{tj} = \Gamma_{ts}$ when $t_s \geq t_i$ and $\Gamma_{tj} = \Gamma_{ti}$ when $t_s < t_i$. $\bar{\Gamma} = (\Gamma_{ts} + \Gamma_{ti})/2$ is the average photon decay rate of the loaded cavity. $g = (c \eta n_2 \hbar \omega_p \sqrt{\omega_s \omega_i})/(n_s n_i \bar V)$ is the vacuum coupling rate for degenerate-SFWM, where c is the speed of light in vacuum, $\eta$ is the spatial overlap fraction between the three modes, $n_2$ is the Kerr nonlinear coefficient, and $n_s$ ($n_i$) is the index of refraction of silicon at the signal (idler) wavelength. Equation (\ref{PairProb}) intuitively demonstrates the benefits of micro/nanoscopic cavity-enhanced SFWM, where high optical Qs ($Q = \omega_0/\Gamma$) and small mode volumes significantly enhance and purify the pair generation process. Moreover, it clearly indicates the strong temporal correlation possessed by photon pairs, a prerequisite for high visibility two-photon interference. 

Franson originally proposed an experiment to observe time-energy entanglement in the context of a three level atomic system, \cite{Franson89} which may be easily extended to the case of photon pairs resulting from SFWM. In the following, we will briefly explore the origin of time-energy entanglement within our system, draw parallels with Franson's proposal, and determine critical requirements for the existence of time-energy entanglement. 

\begin{figure*}[t!]
\begin{center}
\includegraphics[scale=1]{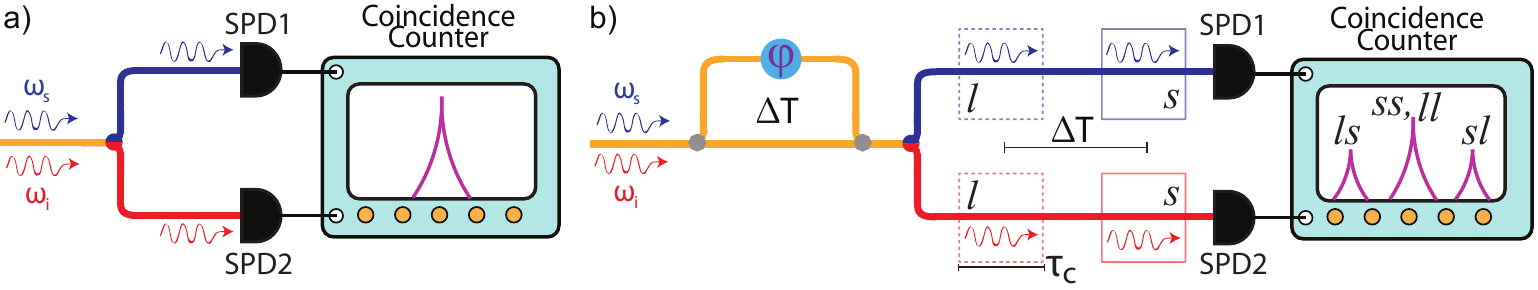}
\caption{ \small Probing the origin of time-energy entanglement. (a) Schematic depicting a conventional coincidence detection setup for analyzing photon pairs generated from SFWM. (b) Schematic depicting the Franson-style detection setup, utilizing a multiplexed unbalanced Mach-Zehnder interferometer (UMZI). The path length difference between the arms of the UMZI results in a time difference, $\Delta T$, and the upper arm of the UMZI has an externally controllable phase, $\phi$. The boxes represent the four pair combinations that result in a coincidence detection. Each pair consists of one photon taken from the signal (blue) channel and another taken from the idler (red) channel. If both photons take the short path (red and blue solid boxes) or the long path (red and blue dashed boxes), they contribute to the central peak. If the signal photon takes the short path (solid blue box) and the idler photon takes the long path (dashed red box), the coincidence adds to the right peak.  If the signal photon takes the long path (dashed blue box) and the idler photon takes the short path (solid red box), the coincidence adds to the left peak. The blue/red circles represent wavelength selective directional couplers (e.g. de-multiplexer), and the gray circles represent 50/50 directional couplers. $\tau_c$ represents the biphoton coherence time. SPD: Single photon detector.}
\label{Fig2}
\end{center}
\end{figure*}

Time-energy entanglement arises naturally when pairs of photons are created through an optical parametric process, such as spontaneous parametric downconversion (SPDC) or SFWM, due to energy conservation \cite{Kwiat93}. The entanglement is consequently a property of the biphoton states, resulting in second-order coherence effects which may be observed in the coincidence detection statistics. To understand why this is the case, consider the two scenarios presented in Figure \ref{Fig2}. Figure \ref{Fig2}(a) depicts a standard coincidence detection setup with photon pairs being split and then detected by single photon detectors (SPD). Here, the signal channel is represented in blue and the idler channel is represented in red. The difference in arrival times between the two channels is recorded with a time-correlated single photon counting (TCSPC) module. As indicated in equation (\ref{PairProb}), the photon pairs exhibit a strong temporal correlation, which results in a single pronounced peak in the coincidence arrival statistics. Figure \ref{Fig2}(b) depicts the same coincidence detection setup, but with the addition of an unbalanced Mach-Zehnder interferometer (UMZI). The introduction of the UMZI causes three peaks to appear in the coincidence spectrum, separated by a time delay, $\Delta T$, defined by the relative path difference between the arms of the interferometer. The boxes in Fig.~\ref{Fig2}(b) represent the four pair combinations that result in a coincidence detection. Each coincidence depends on the arrival time difference between one photon in the signal channel and another photon in the idler channel. If the signal photon takes the long path (dashed blue box) and the idler photon takes the short path (solid red box), the coincidence adds to the left peak. If the signal photon takes the short path (solid blue box) and the idler photon takes the long path (dashed red box), the coincidence adds to the right peak. These paths are clearly distinguishable, due to the difference in arrival times. However, if both photons take the short path (red and blue solid boxes) or if both photons take the long path (red and blue dashed boxes), they contribute to the central peak. Note that although the pairs are created simultaneously, their time of creation is unknown within the long coherence time of the pump laser, $\tau_p$, which precludes distinguishing between short-short and long-long based on an absolute timing reference. Hence, the short-short and long-long arrival times are indistinguishable, which erases the path information, and is thus the source of the quantum interference. To gain further insight regarding the origin of the two-photon interference, consider the energy diagram in Fig.\ref{Fig1}(c). Through conservation of energy, we have the relation that $\omega_s + \omega_i = 2\omega_p$. Furthermore, the pump frequency has an uncertainty, $\Delta\omega_p$, so that $\omega_p = \omega_{p0} + \Delta\omega_p$, where $\omega_{p0}$ is the center frequency. Critically, the uncertainty in the pump frequency is on the order of $\Delta\omega_p \sim 1/\tau_p$, and hence the uncertainty in the sum of the signal and idler frequencies is on the order of $\Delta(\omega_s + \omega_i) \sim 2/\tau_p$. Consequently, the long coherence time of the pump is effectively transferred to the biphotons, so that if $\Delta T \ll \tau_p$, then $\Delta(\omega_s + \omega_i) \Delta T \ll 1$, and the biphoton probability amplitudes at times t and $\rm {t-\Delta T}$ have a fixed phase relationship, as detailed in Franson's paper \cite{Franson89}. Hence, the biphoton states can be made to coherently interfere as a function of UMZI phase, with a theoretical maximum visibility of unity. An additional constraint, $\tau_c \ll \Delta T$, where $\tau_c$ is the biphoton coherence time (see Fig.~\ref{Fig2}(b)), is required to prevent the standard first-order coherence effects. These attributes impose requirements for observing time-energy entanglement, and are summarized by,

\begin{eqnarray}
\tau_c \ll \Delta T \ll \tau_p \label{TimeRelation}
\end{eqnarray}
where, $\tau_c \sim250~{\rm ps}$, $\Delta T \sim16~{\rm ns}$ and $\tau_p \sim2~{\rm \mu s}$, for this experiment, clearly satisfying the requirements of equation (\ref{TimeRelation}). Note that the system depicted in Fig.~\ref{Fig2}(b) involves a single interferometer, which differs from Franson's proposal, wherein the intent is to test the nonlocal description of nature \cite{Franson89,Bell64}. Here we intend to verify and characterize the time-energy entanglement inherent in photon pairs generated from a microdisk device, as well as demonstrate the enhanced phase sensitivity of such states. For these purposes, a single interferometer is sufficient and indeed preferred, due to the ease of implementation and increased system stability. This work aligns with previous investigations of the non-classical nature of light, which have incorporated either a single Michelson interferometer \cite{Martienssen91} or a single Mach-Zehnder interferometer \cite{Mandel90,Rarity90,Ramelow15}.

%%%%%%%%%%%%%%%%%%%%%%%%%%%%%%%%%%%%%%%%%%%%%%%%%%%%%%%%%%%%%%%%%%%%%
%% RESULTS
%%%%%%%%%%%%%%%%%%%%%%%%%%%%%%%%%%%%%%%%%%%%%%%%%%%%%%%%%%%%%%%%%%%%%

\section*{Results}

\begin{figure*}[t!]
\begin{center}
\includegraphics[scale=1]{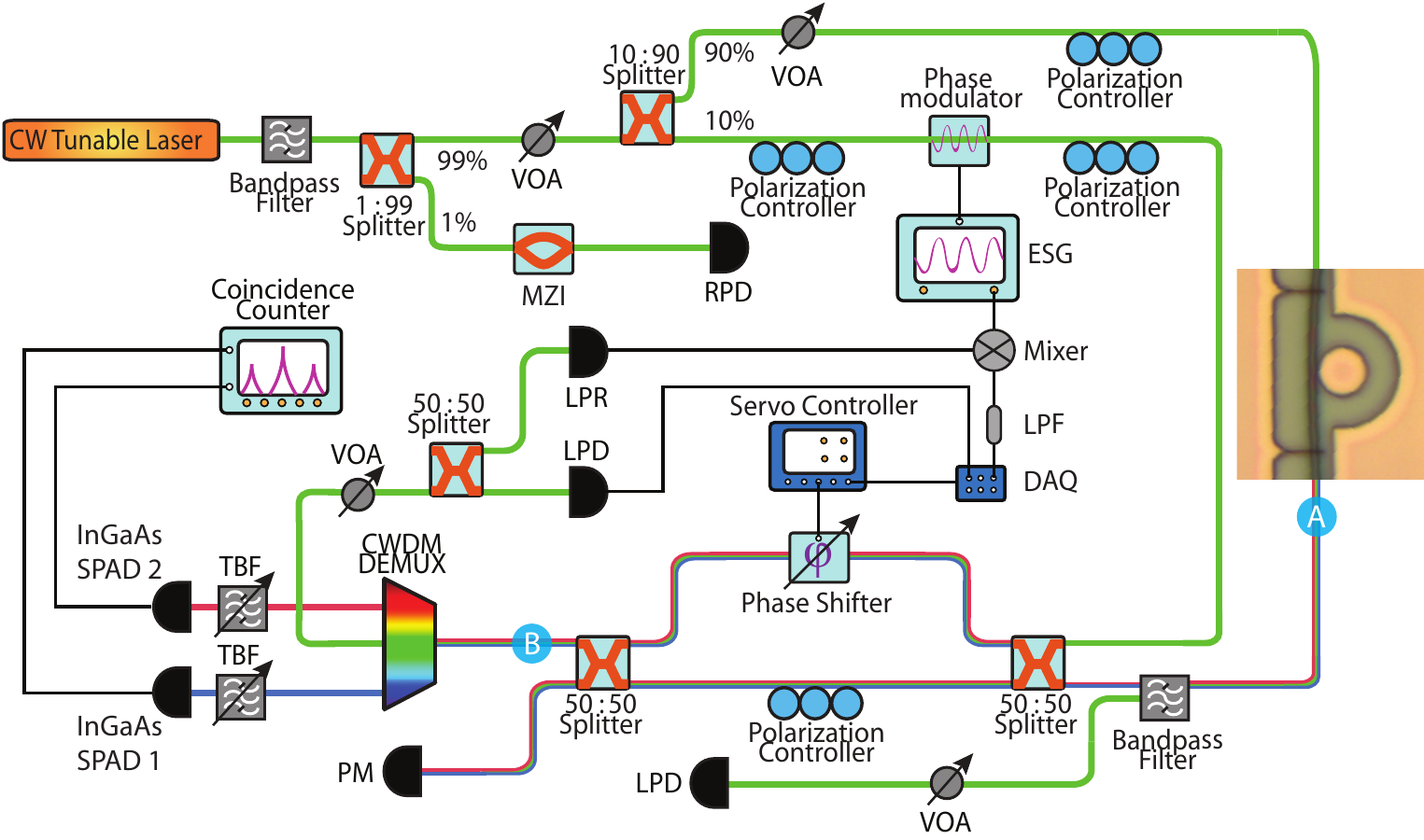}
\caption{ \small Schematic depicting the experimental setup. The experiment includes biphoton generation in a silicon microdisk, the multiplexed unbalanced Mach-Zehnder interferometer (UMZI) for quantum state preparation, the coincidence detection system and a locking technique to stabilize the UMZI. An optical microscopic image of the microdisk is shown in the inset. Note that the colored lines indicate fiber optical pathways, where green, blue and red represent the pump, signal and idler, respectively. Black lines indicate electrical signal pathways. MZI: Mach-Zehnder Interferometer. VOA: Variable Optical Attenuator.  ESG: Electrical Signal Generator. LPF: Low-Pass Filter. DAQ: Data Acquisition. CWDM: Course Wavelength Division Multiplexing. DEMUX: De-Multiplexer. SPAD: Single Photon Avalanche Photodiode. LPD: Locking Photodiode. LPR: Locking Photoreceiver. RPD: Reference Photodiode. TBPF: Tunable Bandpass Filter. Point A indicates the location where the pair generation rate is measured. Point B indicates the location where the entangled pair-flux is measured. }
\label{Fig3}
\end{center}
\end{figure*}

As previously discussed, the presence of time-energy entangled states may be verified by post-selecting the central peak of the signal and idler cross-correlation spectrum. The peak is then expected to exhibit a sinusoidal dependence on the UMZI phase, with a theoretical maximum visibility of unity \cite{Franson89,Kwiat93}. In addition, the two-photon interference - as a result of having a reduced de Broglie wavelength \cite{Yamamoto95,Padua99} - should display twice the fringe density (or half the period) with respect to a classical beam that co-propagates with the biphotons. To probe the time-energy entanglement within our system, an experiment was built and is schematically depicted in Fig.~\ref{Fig3}. The experiment serves three main functions: pair generation through cavity-enhanced SFWM, quantum state preparation with an actively phase stabilized UMZI, and coincidence detection. Each proves critical in observing high coincidence visibility, and the focus of this work pertains to the first two. It is important to note that the quality of the silicon microdisk fundamentally sets an upper bound on the quantum interference visibility, since time-energy entanglement originates from the pair generation process. It is also clear that essentially any application wishing to make use of the unique properties of quantum entanglement requires some form of state preparation in order to exploit this intrinsic quantum nature. Thus, care must be taken with all experimental components contributing to state preparation, as these may impose practical limitations on the achievable interference visibility. In our experiment, it was identified that the UMZI phase stability was critical to realizing high coincidence visibility. In order to achieve this, we developed a convenient phase locking technique that both stabilized the UMZI and allowed for stepping through phase values by adjusting a single system parameter. The entire phase locking scheme was constructed with scalability in mind, so that chip-scale integration is feasible, which would further improve performance by reducing the effects of environmental disturbances. The UMZI allows for the creation of path-entangled states which may be used to verify the non-classical nature of the light. Furthermore, states prepared in this fashion exhibit an increased sensitivity to phase variations, which may find applications in metrology. Additionally, when ultra-pure biphoton sources are used (as is the case here), it is possible to achieve near unity coincidence visibility, but any imperfections in the UMZI may prevent accessing this regime. Hence, the ability to precisely control the UMZI in systems wishing to make use of time-energy entanglement is of extreme importance.  

\begin{figure*}[t!]
\begin{center}
\includegraphics[scale = 1]{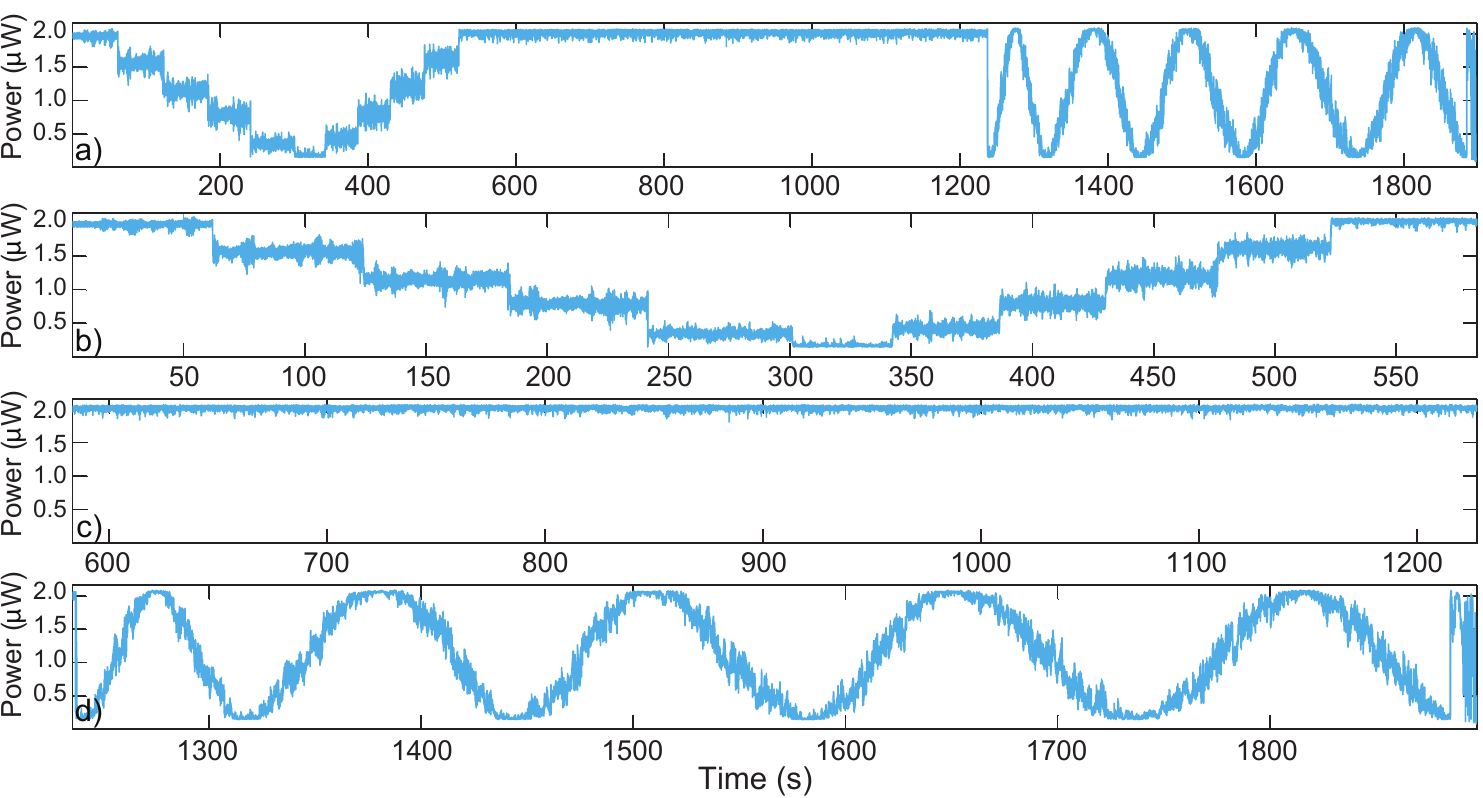}
\caption{ \small Characterization of the active phase stabilization technique. (a) A trace of pump power from the UMZI to characterize the performance of the phase locking scheme under different system settings. (b) Phase values are sequentially locked, resulting in a staircase transmission trace (c) A single phase value is locked for more than 600 seconds. (d) A transmission trace exhibiting significant drift with the locking system disabled.}
\label{Fig4}
%\vspace{-7 mm}
\end{center}
\end{figure*} 

The UMZI phase locking scheme is a variant of the ubiquitous Pound-Drever-Hall laser frequency stabilization technique \cite{Drever83} in conjunction with real-time signal processing. An experimental realization is included as part of the schematic in Fig.~\ref{Fig3}. As seen in Fig.~\ref{Fig3}, a portion of the pump beam is picked off and phase modulated, which creates sidebands with respect to the carrier. The phase modulated pump then propagates through the optical fiber based UMZI, which causes the standard sinusoidal interference pattern at the output. Critically, the sidebands beat with the carrier as all three are swept through the sinusoidal pattern, so that a derivative signal is also produced. This composite signal is then mixed and filtered to isolate the derivative signal. Thus, when the carrier exhibits a sine dependence upon the UMZI phase, the derivative signal will exhibit a cosine dependence with a fixed phase relative to the carrier. The two signals are sent to a data acquisition board (DAQ), so that trigonometric computations may be performed to unwrap the phase of the UMZI, creating a real-time phase reference. The reference is then used as an error signal for the servo controller, which adjusts the UMZI phase shifter and completes the feedback loop. Figure~\ref{Fig4} demonstrates the effectiveness of the phase locking system. Figure~\ref{Fig4}(a) shows the UMZI transmission during a $\sim$~0.5 hour continuous measurement, and Figs.~\ref{Fig4}(b)-(d) are cutouts of the original trace, emphasizing different features. In Fig.~\ref{Fig4}(b), phase values are sequentially locked, which fixes the UMZI transmission in a staircase manner and demonstrates the control of the system. In Fig.~\ref{Fig4}(c), a single phase value is locked for a significant amount of time, indicating the stability of the system. Finally, in Fig.~\ref{Fig4}(d), the phase locking is disabled which allows the UMZI to drift under the influence of external disturbances. Without the locking system, coincidence measurements would depend on a time varying phase, which would destroy the formation of the intended interference pattern. However, with the locking system enabled, phase values may be adjusted over a large range with high resolution. The ease of use and stability of the phase locking scheme make it appealing for a broad range of integrated optical devices, and may be conveniently adapted in most modern optical laboratories. 

\begin{figure*}[t!]
\begin{center}
\includegraphics[scale=1]{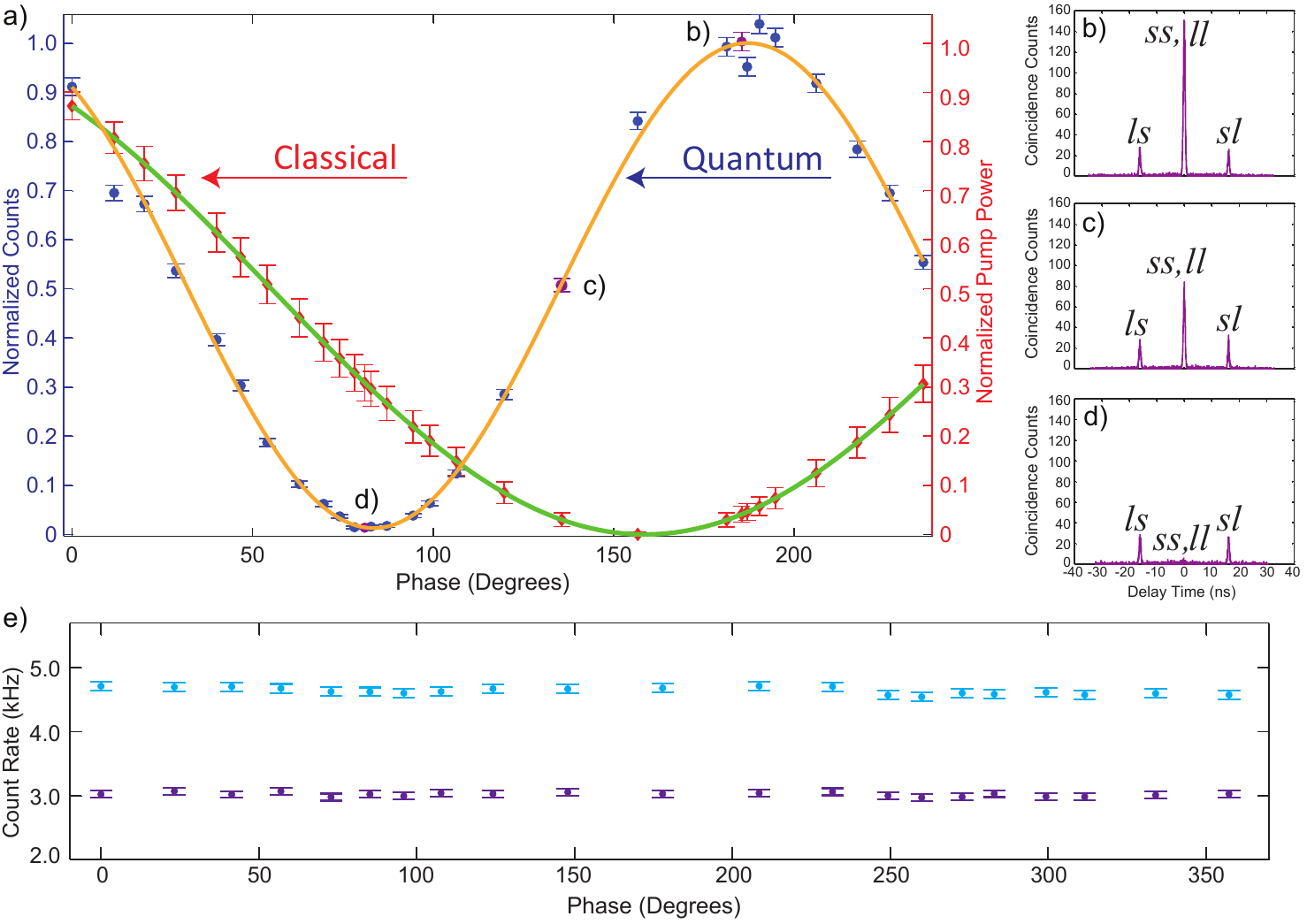}
\caption{ \small A verification of time-energy entanglement through biphoton coincidence visibility. (a) A comparison of classical and quantum interference, with normalized coincidence counts shown in blue and a sinusoidal fit in gold. The normalized pump powers, transmitted through the UMZI, are shown in red with a sinusoidal fit in green. (b)-(d) Raw coincidence spectra with the central peak respectively exhibiting a maximum, midpoint, and minimum. (e) Count rates for the individual channels, with signals shown in purple and idlers shown in cyan.}
\label{Fig5}
\end{center}
\end{figure*}

With the locking system engaged, simultaneous measurements of coincidence counts and pump power from the UMZI were made, for a range of phase values, as shown in Fig.~\ref{Fig5}(a). The normalized coincidence counts, without background subtraction, are shown in blue with a sinusoidal fit in gold. Additionally, the normalized pump powers transmitted through the UMZI are shown in red with a sinusoidal fit in green. Figures~\ref{Fig5}(b)-(d) contain the raw coincidence data corresponding to the labeled points in the coincidence interference pattern. Note that the central peak transforms from a maximum to essentially zero. In particular, the raw coincidence spectra clearly demonstrate that the short-long and long-short states, which produce the side peaks, are unaffected by the UMZI phase, whereas the long-long and short-short states, corresponding to the central peak, exhibit nearly perfect interference. Additionally, the single channel count rates were recorded as the UMZI phase was swept over a full period, as shown in Fig.~\ref{Fig5}(e). The signal channel rates are shown in purple, and the idler channel rates are shown in cyan. It is clear that the single channel rates are independent of UMZI phase, which verifies that first-order coherence effects were not present in the system. 

As shown by the sinusoidal fit in Fig.~\ref{Fig5}(a), the coincidences exhibit a well behaved sinusoidal dependence upon the UMZI phase. A coincidence visibility may be obtained by

\begin{equation}
V_{raw} = \frac{C_{max}-C_{min}}{C_{max}+C_{min}} \label{Vis}
\end{equation}
where $C_{max}$ and $C_{min}$ are the fitted maximum and minimum coincidence counts, respectively. It has been well established that second-order coherence measurements of classical light, involving two indistinguishable two-photon paths, will result in a coincidence visibility of at most 50$\%$ \cite{Kwiat93,Mandel83}. It should be noted that this upper bound may be greatly exceeded if higher-order (beyond second-order) coherence properties are considered and/or the number of indistinguishable two-photon paths exceeds two \cite{Penin08,Zhang13}. The experimental setup depicted in Fig.~\ref{Fig3} precludes these special cases, so that a two-photon interference visibility larger than 50$\%$ requires a non-classical description. By exploiting the unique qualities of cavity-enhanced SFWM in a silicon microdisk, we have observed a raw coincidence visibility of (96.6 $\pm$ 1.1)$\%$, verifying the presence of time-energy entanglement in our system. This remarkably pure quantum interference was achieved with a large pair generation rate of $(4.40\pm0.07) \times 10^5 ~{\rm pairs/s}$. Pair generation rate refers to the pairs of photons delivered from the microdisk into the tapered optical fiber, indicated by point A in Fig.~\ref{Fig3}. It is also important to consider the entangled pair-flux, which is the rate of pairs emerging from the UMZI in a superposition of the short-short and long-long states, indicated by point B in Fig.~\ref{Fig3}. The entangled pair-flux was measured to be $(3.10 \pm 0.05) \times 10^4 ~{\rm pairs/s}$. Moreover, the high spectral brightness of the device allowed for the experiment to be performed at an ultra-low optical power dropped into the cavity of $P_d = 22~\mu W$. Additionally, the biphoton interference is expected to exhibit half the period of the interference pattern generated by a classical beam through the same interferometer. Fits to the data in Fig.~\ref{Fig5}(a) reveal that the ratio of classical to quantum period is 1.996, agreeing well with theory \cite{Yamamoto95,Padua99} and is a verification of the enhanced sensitivity of the quantum light. 

%%%%%%%%%%%%%%%%%%%%%%%%%%%%%%%%%%%%%%%%%%%%%%%%%%%%%%%%%%%%%%%%%%%%%
%% DISCUSSION
%%%%%%%%%%%%%%%%%%%%%%%%%%%%%%%%%%%%%%%%%%%%%%%%%%%%%%%%%%%%%%%%%%%%%

\section*{Discussion}
Photon pairs generated from high-Q silicon microdisk resonators have been shown to exhibit remarkable qualities that align well with quantum photonic platforms. We have extended this domain, by verifying the first realization of time-energy entangled photon pairs from a microdisk device. The combination of an excellent photon source and robust phase locking technique, enabled the observation of a two-photon interference visibility of (96.6 $\pm$ 1.1)$\%$, without any background subtraction, at a pair generation rate of $(4.40\pm0.07) \times 10^5 ~{\rm pairs/s}$ and ultra-low pump power. To the best of our knowledge, this is the highest raw quantum interference visibility for time-energy entangled photon pairs from a micro/nanoscale source, to date  \cite{Ramelow15,Obrien13,Wakabayashi15,Preble15,Grassani15,Xiong15}. Furthermore, a novel pump phase modulation scheme was presented, which allowed for convenient stabilization and control of the UMZI phase. Phase locking techniques of this sort may find usefulness in interferometric quantum state preparation, quantum metrology setups, as well as classical photonics applications. 

As lab-on-a-chip technology moves closer to reality for quantum optics, it is important to consider the vital role assumed by photon sources in such a system, as well as their relationship to other integrated components. Recent works have demonstrated that quantum gate operations of essentially unity fidelity are achievable using integrated photonic circuits \cite{OBrien10,EnglundSep15,Tang16}. Additionally, waveguide integrated superconducting nanowire single photon detectors (SNSPD) of high detection efficiency \cite{Pernice15,Young15} and integrated networks of SNSPDs with significant system detection efficiency \cite{EnglundJan15} have also been realized. These advances further motivate the importance of chip-scale sources of highly pure photons, which impose fundamental limits on the fidelity in integrated quantum photonic processes. High-Q silicon microdisks, fabricated using the CMOS compatible platform and designed to emit pairs of photons at the important telecommunication wavelength band show great promise for future quantum photonics applications.  

%%%%%%%%%%%%%%%%%%%%%%%%%%%%%%%%%%%%%%%%%%%%%%%%%%%%%%%%%%%%%%%%%%%%%
%% Acknowledgments
%%%%%%%%%%%%%%%%%%%%%%%%%%%%%%%%%%%%%%%%%%%%%%%%%%%%%%%%%%%%%%%%%%%%%
\begin{acknowledgments}
This work was supported by the National Science Foundation under Grant No.~ECCS-1351697. It was performed in part at the Cornell NanoScale Facility, a member of the National Nanotechnology Coordinated Infrastructure (NNCI), which is supported by the National Science Foundation (Grant ECCS-1542081). 
\end{acknowledgments}

%%%%%%%%%%%%%%%%%%%%%%%%%%%%%%%%%%%%%%%%%%%%%%%%%%%%%%%%%%%%%%%%%%%%%
%% APPENDIX
%%%%%%%%%%%%%%%%%%%%%%%%%%%%%%%%%%%%%%%%%%%%%%%%%%%%%%%%%%%%%%%%%%%%%

\section*{Appendix}
\subsection*{Experimental setup and device characterization}

The schematic depicted in Fig.~\ref{Fig3} details all of the essential components of the experiment. A tunable continuous-wave laser (linewidth $\sim200$~kHz) is coupled into an optical fiber and passed through the transmission port of a bandpass filter (BPF). The BPF has a 3-dB bandwidth of 17 nm and band isolation  exceeding 120 dB, which serves to suppress the amplified spontaneous emission (ASE) from the pump laser. This is necessary due to the broadband nature of ASE, which would otherwise corrupt the single photon measurements performed on the signal and idler channels. The pump light is then coupled into the microdisk through a tapered optical fiber and controlled by a nano-positioning setup. Real-time monitoring of the cavity transmission allows for desired coupling conditions to be achieved and fixed by anchoring the fiber against the device nanoforks, as seen in the optical microscopic image in Fig.~\ref{Fig3}. Variable optical attenuators (VOA) are used to control the pump power delivered to the device and fiber polarization controllers (FPC) are used to align the pump polarization state to that of the cavity resonances (quasi-TM for this experiment). The pump and biphotons exit the cavity in the same optical fiber and are subsequently isolated by a BPF. The biphotons then pass through the UMZI and onto a course wavelength division multiplexing (CWDM) de-multiplexer (DEMUX), which separates the signal and idler channels. These channels propagate through tunable bandpass filters (TBPF), having a 3-dB bandwidth of 1.2 nm. The TBPFs are used to suppress Raman noise photons from the optical fiber. It was experimentally verified that Raman noise photons originate from the optical fiber, but not the device \cite{Jiang15}. The biphotons are detected using two gated InGaAs single photon avalanche photodiodes (SPAD), operated in Geiger-mode and thermo-electrically cooled. 

The dispersion measurements, presented in the inset of Fig.~\ref{Fig1}(b), were made using a calibrated wavemeter. The cavity transmission trace (see Fig.~\ref{Fig1}(b)) was acquired by sweeping the wavelength of the pump laser (with BPFs bypassed) while recording the transmitted light on a fast photodiode. A small portion of the pump is coupled directly to a Mach-Zehnder interferometer (MZI), which calibrates the wavelength and is used for fitting the optical resonances. The photoluminescence spectrum (see Fig.~\ref{Fig1}(d)) was measured directly from the DEMUX, using a liquid nitrogen cooled spectrometer.

\subsection*{Coincidence detection, pair generation rate, and entangled pair-flux}

Coincidence detection was performed with both SPADs set to a gating frequency of 2.5 MHz, gate width of 40 ns, dead time of 10 $\rm {\mu s}$, and quantum efficiency of 15$\%$. The total coincidence counts, $C_t$, were obtained by integrating the central peak of the cross-correlation spectra over a coincidence window, $\Delta\tau = \tau_{\rm FWHM}$. $\tau_{\rm FWHM}$ was established by the full-width-half-maximum (FWHM) of the coincidence spectrum with the largest central peak. The statistical error for the total coincidence counts is given by $E_{C_t} = \sqrt{C_t}$. The accidental counts, $A_t$, were obtained by integrating the noise floor adjacent to the central peak, using the same coincidence window. 

The pair generation rate was obtained by calibrating the coincidence counts per gate with the clock frequency, duty cycle, and quantum efficiency of the detectors along with the transmittance from the microdisk to the SPADs. The transmittance for the signal and idler were $\eta_{\rm s}$ = 0.37 and $\eta_{\rm i}$ = 0.35, respectively. The transmittance includes the taper loss and insertion loss of all components from the device (see point A in Fig.~\ref{Fig3}) to the SPADs. 

The entangled pair-flux was obtained in the same manner as the pair generation rate, but for pairs exiting the UMZI (see point B in Fig.~\ref{Fig3}) instead of the microdisk. Here, the transmittance for the signal and idler were $\eta_{\rm s}$ = 0.56 and $\eta_{\rm i}$ = 0.52, respectively.

%%%%%%%%%%%%%%%%%%%%%%%%%%%%%%%%%%%%%%%%%%%%%%%%%%%%%%%%%%%%%%%%%%%%%
%% REFERENCES
%%%%%%%%%%%%%%%%%%%%%%%%%%%%%%%%%%%%%%%%%%%%%%%%%%%%%%%%%%%%%%%%%%%%%


\begin{thebibliography}{99}

%Quantum communication
\bibitem{Gisin07} %1
	Gisin, N. \& Thew, R. Quantum communication. \emph{Nat. Photonics} {\bf 1,} 165 (2007).	
\bibitem{Zbinden02} %2    
    Gisin, N., Ribordy, G., Tittel, W., \& Zbinden, H. Quantum cryptography. \emph{Rev. Mod. Phys.} {\bf 74,} 145-195 (2002).
\bibitem{Villoresi15} %3
	Vallone, G. \emph{et al.} Experimental satellite quantum communications. \emph{Phys. Rev. Lett.} {\bf 115,} 040502 (2015).  

%Quantum simulators	
\bibitem{Aspuru-Guzik12} %4 
	Aspuru-Guzik, A. \&  Walther, P. Photonic quantum simulators. \emph{Nature Phys.} {\bf 8,} 285 (2012).
\bibitem{Pan09} %5
	Lu, C.-Y. et al. Demonstrating anyonic fractional statistics with a six-qubit quantum simulator \emph{Phys. Rev. Lett.} {\bf 102,} 030502 (2009).

%Quantum computing
\bibitem{Milburn01} %6
	Knill, E., Laflamme, R., \& Milburn, G. J. A scheme for efficient quantum computation with linear optics. \emph{Nature} {\bf 409,} 46–52 (2001).
\bibitem{Milburn07} %7    
    Kok, P. \emph{et al.} Linear optical quantum computing with photonic qubits. \emph{Rev. Mod. Phys.} {\bf 79,} 135-174 (2007).
\bibitem{Pittman02} %8
	Pittman, T. B., Jacobs, B. C., \& Franson, J. D. Demonstration of nondeterministic quantum logic operations using linear optical elements. \emph{Phys. Rev. Lett.} {\bf 88,} 257902 (2002).
\bibitem{OBrien09} %9
	O'Brien, J. L., Furusawa, A., \& Vuckovic, J. Photonic quantum technologies. \emph{Nat. Photonics} {\bf 3,} 687-695 (2009).
\bibitem{OBrien15} %10
	Carolan, J. \emph{et al.} Universal linear optics. \emph{Science} {\bf 349,} 711 (2015).

%Quantum metrology
\bibitem{Dowling02} %11
	Lee, H., Kok, P., \& Dowling, J. P. A quantum Rosetta stone for interferometry. \emph{J. Mod. Opt.} {\bf 49,} 2325–2338 (2002). 
\bibitem{Maccone04} %12
	Giovannetti, V., Lloyd, S., \& Maccone, L. Quantum-enhanced measurements: Beating the standard quantum limit. \emph{Science} {\bf 306,} 1330-1336 (2004). 

%High visibility quantum interference for q. photonic applications
\bibitem{Zeilinger12} %13
	Pan, J.-W. et al. Multiphoton entanglement and interferometry. \emph{Rev. Mod. Phys.} {\bf 84,} 777-838 (2012).
	
%Microscale silicon based sources of time-energy entangled photons
\bibitem{Obrien13} %14
	Silverstone, J. W. \emph{et al.} On-chip quantum interference between silicon photon-pair sources. \emph{Nat. Photonics} {\bf 8,} 104 (2013).
\bibitem{Wakabayashi15} %15
	Wakabayashi, R. \emph{et al.} Time-bin entangled photon pair generation from Si micro-ring resonator. \emph{Opt. Express} {\bf 23,} 1103-1113 (2015).
\bibitem{Preble15} %16
	Preble, S. F. \emph{et al.} On-chip quantum interference from a single silicon ring resonator source. \emph{Phys. Rev. Applied} {\bf 4,} 021001 (2015). 
\bibitem{Grassani15} %17
	Grassani, D. \emph{et al.} Micrometer-scale integrated silicon source of time-energy entangled photons. \emph{Optica} {\bf 2,} 88-94 (2015).
\bibitem{Xiong15} %18
	Xiong, C. \emph{et al.} Compact and reconfigurable silicon nitride time-bin entanglement circuit. \emph{Optica} {\bf 2,} 724-727 (2015).
\bibitem{Ramelow15} %19
	Ramelow, S. \emph{et al.} Silicon-nitride platform for narrowband entangled photon generation.  arXiv:1508.04358v1 (2015).
	
%High-Q silicon microdisk resonators 
\bibitem{Jiang15}   %20 
    Jiang, W. C., Lu, X., Zhang, J., Painter, O., \& Lin, Q. Silicon-chip source of bright photon pairs. \emph{Opt. Express} {\bf 23,} 20884-20904 (2015).
\bibitem{Lu15} %21
    Lu, X., Jiang, W. C., Zhang, J., \& Lin, Q. Biphoton statistics of quantum light generated on a silicon chip. arXiv:1602.08057v2 (2016).
\bibitem{Rogers15}  %22 
    Rogers, S., Lu, X., Jiang, W. C., \& Lin, Q. Twin photon pairs in a high-Q silicon microresonator. \emph{Appl. Phys. Lett.} {\bf 107,} 041102 (2015).
    
%SFWM process
\bibitem{Boyd08} %23 
    Boyd, R. W. \emph{Nonlinear Optics}, $3^{\rm rd}$ Ed. (Academic Press, New York, 2008).	
    
%Franson
\bibitem{Franson89} %24
	Franson, J. D. Bell inequality for position and time. \emph{Phys. Rev. Lett.} {\bf 62,} 2205-2208 (1989).
	
%Origin of time-energy entanglement
\bibitem{Kwiat93} %25
	Kwiat, P.G., Steinberg, A. M., \& Chiao, R. Y. High-visibility interference in a Bell-inequality experiment for energy and time. \emph{Phys. Rev. A} {\bf 47,} R2472 (1993). 

%Nonclassical verification of light in a single Michelson interferometer
\bibitem{Martienssen91} %26
	Brendel, J., Mohler, E., \& Martienssen, W. Time-resolved dual-beam two-photon interferences with high visibility. \emph{Phys. Rev. Lett.} {\bf 66,} 1142-1145 (1991).

%Nonclassical two-photon interference in a single Mach-Zehnder interferometer
\bibitem{Mandel90} %27
	Ou, Z. Y., Zou, X. Y., Wang, L. J., \& Mandel, L. Experiment on nonclassical fourth-order interference. \emph{Phys. Rev. A} {\bf 42,} 2957-2965 (1990).
\bibitem{Rarity90} %28
	Rarity, J. G. \emph{et al.} Two-photon interference in a Mach-Zehnder interferometer. \emph{Phys. Rev. Lett.} {\bf 65,} 1348-1351 (1990).


%Bell inequality
\bibitem{Bell64} %29
	Bell, J.S. On the Einstein Podolsky Rosen paradox. \emph{Physics} {\bf 1,} 195-200 (1964).

%Reduced photonic de Broglie wavelength
\bibitem{Yamamoto95} %30
	Jacobsen, J., Bjork, G., Chuang, I., \& Yamamoto, Y. Photonic de Broglie waves. \emph{Phys. Rev. Lett.} {\bf 74,} 4835-4838 (1995).
\bibitem{Padua99} %31
	Fonseca, E. J. S., Monken, C. H., \& Padua, S. Measurement of the de Broglie wavelength of a multiphoton wave packet. \emph{Phys. Rev. Lett.} {\bf 82,} 2868-2871 (1999).
	
%PDH laser frequency stabilization
\bibitem{Drever83} %32
	Drever, R. W. P. \emph{et al.} Laser phase and frequency stabilization using an optical resonator. \emph{Appl. Phys. B: Photophys. Laser Chem.} {\bf 31}, 97-105 (1983).

%Upper bound for classical light two photon interference
\bibitem{Mandel83} %33
	Mandel, L. Photon interference and correlation effects produced by independent quantum sources. \emph{Phys. Rev. A} {\bf 28} 929-943 (1983).


%Greater than 50% classical vis for higher-order correlations or more than two two-photon paths
\bibitem{Penin08} %34
	Agafonov, I. N., Chekhova, M. V., Iskhahov, T. Sh., \& Penin, A. N. High-visibility multiphoton interference of Hanbury Brown-Twiss type for classical light. \emph{Phys. Rev. A} {\bf 77,} 053801 (2008).
\bibitem{Zhang13} %35
	Hong, P., Xu, L., Zhai, Z., \& Zhang, G. High visibility two-photon interference with classical light. \emph{Opt. Express} {\bf 21,} 14056-14065 (2013).
	
%High fidelity quantum gate operations
\bibitem{OBrien10} %36
	Laing, A. \emph{et al.} High-fidelity operation of quantum photonic circuits. \emph{Appl. Phys. Lett.} {\bf 97,} 211109 (2010).
\bibitem{EnglundSep15} %37
	Mower, J., Harris, N. C., Steinbrecher, G. R., Lahini, Y., \& Englund, D. High-fidelity quantum state evolution in imperfect photonic integrated circuits. \emph{Phys. Rev. A} {\bf 92,} 032322 (2015).
\bibitem{Tang16} %38
	Poot, M., Schuck, C., Ma, X., Guo, X., \& Tang, H. X. Design and characterization of integrated components for SiN photonic quantum circuits. \emph{Opt. Express} {\bf 24,} 6843-6860 (2016).
	
%High detection efficiency waveguide integrated SNSPDs
\bibitem{Pernice15} %39
	Kahl, O. \emph{et al.} Waveguide integrated superconducting single-photon detectors with high internal quantum efficiency at telecom wavelengths. \emph{Sci. Rep.} {\bf 5,} 10941 (2015).
\bibitem{Young15} %40
	Akhlaghi, M. K., Schelew, E., \& Young, J. F. Waveguide integrated superconducting single-photon detectors implemented as near-perfect absorbers of coherent radiation. \emph{Nat. Commun.} {\bf 6,} 8233 (2015).
	
%Scalable integration of SNSPDs
\bibitem{EnglundJan15} %41
	Najafi, F. \emph{et al.} On-chip detection of non-classical light by scalable integration of single-photon detectors. \emph{Nat. Commun.} {\bf 6,} 5873 (2015).


\end{thebibliography}
\end{document}